\documentclass[prd,amsfonts,noshowpacs,nofootinbib]{revtex4}
\usepackage{amsmath,graphicx}
 
\newcommand{\eq}{\!\! =\!\!}

\def\k{{\bf k}}

\begin{document}
\title{Trispectrum estimation
in various models of equilateral type non-Gaussianity}
\author{Keisuke Izumi\footnote{izumi@phys.ntu.edu.tw}$\sharp$}
\author{Shuntaro Mizuno
\footnote{shuntaro.mizuno@th.u-psud.fr}$\natural$,$\flat$}
\author{Kazuya Koyama\footnote{Kazuya.Koyama@port.ac.uk}$\|$}
\affiliation{$\sharp$
Leung Center for Cosmology and Particle Astrophysics,
National Taiwan University, Taipei 10617, Taiwan, R.O.C.}
\affiliation{$\natural$ Laboratoire de Physique Th\'eorique,
Universit\'e Paris-Sud 11 et CNRS, B\^atiment 210, 91405 Orsay
Cedex, France
}
\affiliation{$\flat$ APC (CNRS-Universit\'e Paris 7),
10 rue Alice Domon et L\'eonie Duquet, 75205 Paris Cedex 13,
France}
\affiliation{$\|$ Institute of Cosmology and Gravitation, University of Portsmouth, Portsmouth PO1 3FX, UK.
}

\date{\today}
\begin{abstract}
We calculate the shape correlations
between trispectra in various equilateral non-Gaussian models, including
DBI inflation, ghost inflation and Lifshitz scalars, 
using the full trispectrum as well as the reduced trispectum. 
We find that most theoretical models are distinguishable
from the shapes of primordial trispectra 
except for several exceptions where it is 
difficult to discriminate between the models,
such as single field DBI inflation and a Lifshitz scalar model.
We introduce an estimator for the amplitude of the trispectrum, 
$g_{\rm NL} ^{equil}$ and relate it to model parameters in various 
models. Using constraints on $g_{\rm NL} ^{equil}$ from WMAP5, we 
give constraints on the model parameters.
\end{abstract}

\maketitle

\section{Introduction}

Almost scale-invariant and Gaussian primordial cosmological perturbations
predicted by inflation are consistent with observational data.
While this suggests that inflation did happen in the early universe,
there still remain many important questions about inflation.
One of the most important problems is to distinguish between
various inflationary models as there are still many models that are consistent
with observational data at present.
For this purpose, it is necessary to have further
information about the early universe such as deviations
from Gaussianity (non-Gaussianity) of the primordial perturbations
and primordial gravitational waves. In this paper, we focus on primordial
non-Gaussianity.

Actually, the statistical properties of primordial fluctuations provide crucial
information on the physics of the very early universe
\cite{Komatsu:2010fb, Smith:2009jr,
Senatore:2009gt, Fergusson:2010dm}
(see \cite{Bartolo:2004if} for a review).
In the simplest single field inflation models where the scalar field has a canonical
kinetic term and quantum fluctuations are generated from the standard
Bunch-Davis vacuum, non-Gaussianity of the fluctuations is too small to
be observed even with future experiments
\cite{Acquaviva:2002ud, Maldacena:2002vr, Seery:2005wm}.
Thus, the detection of
non-negligible deviations from Gaussianity of primordial fluctuations
will have a huge impact on the models of the early universe.
So far, most of the studies have focused on the leading order non-Gaussianity measured
by the three-point function of Cosmic Microwave Background
(CMB) anisotropies,
i.e. the bispectrum
\cite{Verde:1999ij,Wang:1999vf,Komatsu:2001rj}.
Especially, the optimal method of extracting
the bispectrum from the CMB data has been sufficiently
developed
\cite{Komatsu:2003iq, Babich:2004gb, Babich:2005en, Creminelli:2005hu,
Creminelli:2006gc, Yadav:2007rk, Yadav:2007ny}
(for a more general approach, see
\cite{Fergusson:2006pr, Fergusson:2008ra,Fergusson:2009nv}).
However, the bispectrum includes only a part of
information about
non-Gaussianity and many models can predict similar
bispectra.

For example, k-inflation
\cite{ArmendarizPicon:1999rj,Garriga:1999vw}
and Dirac-Born-Infeld (DBI) inflation \cite{Silverstein:2003hf}
are shown to predict almost equilateral type bispectrum
(see \cite{Koyama:2010xj, Chen:2010xka} for reviews).
However, future experiments
like Planck \cite{Planck} can also prove the higher order statistics
such as the trispectrum
\cite{Hu:2001fa,Okamoto:2002ik,Kogo:2006kh}
which gives information that cannot be obtained from the
bispectrum \cite{Seery:2006js, Seery:2006vu, Byrnes:2006vq}.

Non-Gaussianity of the observed CMB anisotropies comes from not only the primordial origin
but also the non-linear effects in the CMB at late times such as the coupling between
Integrated Sachs-Wolfe (ISW) effect and the weak gravitational lensing~\cite{arXiv:1003.6097}.
However, these non-linear effects can be negligible compared with the primordial non-Gaussainity
in the models such as DBI inflation and ghost inflation where large primordial non-Gaussianity
is predicted and it is the dominant contribution to the observed non-Gaussianity.
In this paper, we focus on large primordial non-Gaussianity.

While the trispectrum has more information, it requires more work to understand how to distinguish between various trispectra predicted in many theoretical models and
how to measure the amplitude of the trispectrum because it has more parameters than
the bispectrum. To estimate an overlap between
the shapes of two different trispectra,
the shape correlator was
introduced by Regan et.al \cite{2010arXiv1004.2915R}
based on the reduced trispectrum. Furthermore, based on this
shape correlator, two of us investigated an estimator
$g_{\rm NL} ^{equil}$ to measure the amplitude of the trispectrum
in some equilateral type non-Gaussian models, like
k-inflation, single field DBI inflation and multi-field DBI
inflation \cite{Mizuno:2010by}.

Recently, the shape dependence of
the trispectra from ghost inflation
model~\cite{ArkaniHamed:2003uy,ArkaniHamed:2003uz} and
Lifshitz scalar model have been calculated in Refs.~\cite{Izumi:2010wm, Huang:2010ab}
and Ref.~\cite{Izumi:2010yn}.  These models are known to
give equilateral type bispectra.
For Lifshitz scalar model, we will see that
there is no natural way to decompose the full trispectrum
into the reduced trispectra in some cases 
and the shape correlator
based on the reduced trispectrum is not necessarily
well-defined.
Therefore, in this paper,
we consider an implementation of the shape correlator
of the primordial trispectrum using the full
trispectrum so that we can calculate the shape correlations
and investigate estimators in equilateral type non-Gaussian
models including ghost inflation and Lifshitz scalar.

The rest of this paper is organised as follows.
In section~\ref{shapecorrelator}, we introduce two
shape correlators defined in different ways
where one is defined based on the reduced trispectrum
and the other is based on the full trispectrum.
In section~\ref{Shapecorrelations}, we calculate the shape correlations
of the trispectra among various theoretical models;
DBI inflation, ghost inflation and Lifshitz scalar.
In section~IV, we express the estimators
$g_{\rm NL} ^{equil}$ in terms of the
model parameters, which is useful to constrain
the model parameters from future experiments.
We also give constraints on model parameters using WMAP 
constraints on $g_{\rm NL}^{equil}$ obtained in \cite{Fergusson:2010gn}.
Section~\ref{sec:summary}
is devoted to the summary and discussions
of this paper.
In appendix sections~\ref{appDBI}, \ref{appg} and \ref{apph},
we review the shape functions of the reduced trispectra
in DBI inflation model, ghost inflation model
and Lifshitz scalar model.

\section{Two types of shape correlators}
\label{shapecorrelator}

In this section, we introduce two shape correlators.
One is introduced by Regan~\cite{2010arXiv1004.2915R}
and we review this method in section~\ref{Regan}.
In this method, the shape correlators are defined based on
the reduced trispectrum.
Although the reduced trispectum includes all information of the original trispectrum, there is no unique way to decompose the total trispectrum into the reduced trispectrum.
Then there appears an ambiguity in the definition of shape correlator based on the reduced trispectrum; a different
choice of the reduced trispectrum gives a different shape correlator.

Before discussing the shape correlators,
we review the definition of the trispectrum.
The trispectrum $T_{\zeta}(\k_1,\k_2,\k_3,\k_4)$ of the curvature perturbation
$\zeta$ is
defined as
\begin{equation}
\langle \zeta(\k_1) \zeta(\k_2) \zeta(\k_3) \zeta(\k_4) \rangle_c
=(2 \pi)^3 \delta^3(\k_1+\k_2+\k_3+\k_4)
T_{\zeta}(\k_1,\k_2,\k_3,\k_4),
\label{deftri}
\end{equation}
where $\zeta(\k_1)$ is a Fourier component with the momentum $\k_1$ and
the subscript ``$c$" in the left hand side denotes the connected component.
The trispectrum $T_{\zeta}(\k_1,\k_2,\k_3,\k_4)$ generally depends on four
three-momenta, namely twelve parameters.
Assuming isotropy and homogeneity of the universe on
large scales, the number of parameters reduces to six.

\subsection{The shape correlator based on the reduced trispectrum}
\label{Regan}
Here, we review the shape correlator discussed in  \cite{2010arXiv1004.2915R}.
First we exploit the symmetry of the trispectrum
to define the reduced trispectrum as follows~\cite{Hu:2001fa}.
We rewrite the definition of the trispectrum as
\begin{eqnarray}
\langle \zeta({\bf k_1})  \zeta({\bf k_2})
 \zeta({\bf k_3})  \zeta({\bf k_4})\rangle_c
&=&(2 \pi)^3 \int d^3 K \bigl[
\delta({\bf k_1} + {\bf k_2} - {\bf K})
\delta({\bf k_3} + {\bf k_4} + {\bf K})
\mathcal{T}_\zeta ({\bf k_1}, {\bf k_2}, {\bf k_3}, {\bf k_4};
{\bf K}) \nonumber \\
&& + ({\bf k_2} \leftrightarrow {\bf k_3}) +
({\bf k_2} \leftrightarrow {\bf k_4})
\bigr]\,.
\label{defrtri}
\end{eqnarray}
Because of the symmetry of the trispectrum,
the reduced trispectrum includes all information of the original
trispectrum~\cite{Hu:2001fa}.
However, this decomposition of the trispectrum is not unique
because there is an ambiguity in choosing {{\bf K}} in Eq.~(\ref{defrtri}).
In some cases, there is a natural choice of the reduced trispectrum.
For example, in the case of trispectrum produced by
two three-point vertices, it consists of  s-, t- and u-channels. Then
we can decompose the trispectrum into three parts accordingly and define
the reduced trispectrum in a natural way.

 \begin{figure}[t]
 \begin{center}
    \includegraphics[keepaspectratio=true,height=50mm]{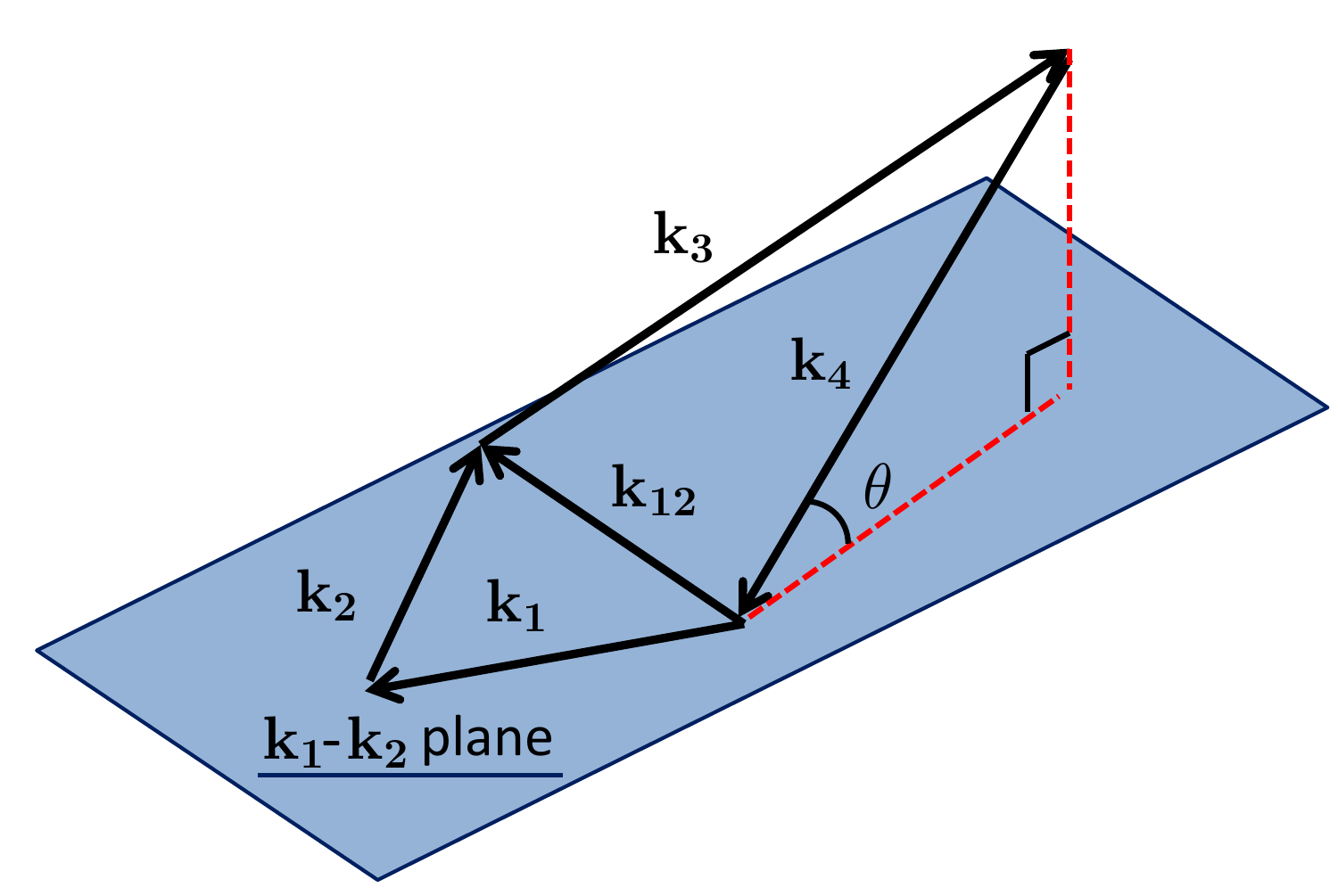}
  \end{center}
  \caption{definition of $\theta_4$}
  \label{fig:theta.eps}
\end{figure}

The reduced trispectrum
$\mathcal{T}_\zeta ({\bf k_1}, {\bf k_2}, {\bf k_3}, {\bf k_4};{\bf K})$
depends on six variables and
we can choose them to be $(k_1,k_2,k_3,k_4,k_{12},\theta_4)$
where $\theta_4$ represents the deviation of the quadrilateral from planarity which is specified by the triangle $(k_1, k_2, k_{12})$ (see Fig~\ref{fig:theta.eps}).
From the geometric restriction, the range of $\theta_4$ is constrained
as
\begin{eqnarray}
 \frac{|k_4^2 + k_{12} ^2 -k_3 ^2|}{2 k_{12} k_4} \leq \cos \theta_4 \leq 1.
\label{const_costheta4}
\end{eqnarray}

Motivated by the relation between the CMB trispectrum and the trispectrum for $\zeta$, we define the shape function for the reduced trispectrum as
\footnote{The factor to relate the reduced trispectrum with
the shape function is not unique. For example,
in Ref.~\cite{2010arXiv1004.2915R}, instead of
$(k_1 k_2 k_3 k_4)^2 k_{12}$, another choice
$(k_1 k_2 k_3 k_4)^{9/4}$ is also discussed.
But it is possible to check that the dependence on this factor
is not very significant when the shape correlation
is sufficiently large, like $\bar{\mathcal{C}} (S_{\mathcal{T}},
S_{\mathcal{T}'}) > 0.7$.}
\begin{eqnarray}
S_{\mathcal{T}}(k_1, k_2, k_3, k_4, k_{12}, \theta_4)
= (k_1 k_2 k_3 k_4)^2 k_{12}
\mathcal{T}_\zeta (k_1, k_2, k_3, k_4; k_{12}, \theta_4)\,.
\label{shapefunc_def}
\end{eqnarray}
Regan et al. \cite{2010arXiv1004.2915R} proposed to define an
inner product between
two different shape functions $S_{\mathcal{T}}$
and $S_{\mathcal{T}'}$ as
\begin{eqnarray}
&&F(S_{\mathcal{T}}, S_{\mathcal{T}'}) \nonumber\\
&&\qquad
= \int d k_1 d k_2 d k_3 d k_4 d k_{12} \int d (\cos \theta_4)
S_{\mathcal{T}} (k_1, k_2, k_3, k_4, k_{12}, \theta_4)
S_{\mathcal{T}'} (k_1, k_2, k_3, k_4, k_{12}, \theta_4)
w(k_1, k_2, k_3, k_4, k_{12})\,,
\label{pritrispectrum_correlator_part}
\end{eqnarray}
where $w$ is an appropriate weight function.
The weight function should be chosen
such that $S^2 w$ in $k$ space produces the same scaling
as the estimator in $l$ space
and we adopt the one used in Ref.~\cite{2010arXiv1004.2915R}
\footnote{The choice of the weight function
is not unique, either. For example,
in the first version of Ref.~\cite{2010arXiv1004.2915R},
instead of
$\frac{k_{12}}{ (k_1 + k_2 + k_{12})^2(k_3 + k_4 + k_{12})^2}$,
another choice
$\frac{1}{k_{12} (k_1 + k_2 + k_{12})(k_3 + k_4 + k_{12})}$
is considered.
But again, we checked that
the dependence on this factor is not significant when the shape correlation
is sufficiently large.},
\begin{eqnarray}
w (k_1, k_2, k_3, k_4, k_{12})
= \frac{k_{12}}{ (k_1 + k_2 + k_{12})^2(k_3 + k_4 + k_{12})^2}\,.
\label{trispectrum_weight}
\end{eqnarray}
The integration range of the momenta $k_1$, $k_2$, $k_3,$ $k_4$, $k_{12}$
in the integral of Eq.~(\ref{pritrispectrum_correlator_part})
is determined by the triangle inequality of the momenta.
Namely, the conditions to satisfy ${\bf k_{12}}={\bf k_1} + {\bf k_2}$ are
\begin{eqnarray}
k_{12}\le k_1 + k_2, \qquad
k_1\le k_{12} + k_2 \qquad \mbox{and} \qquad
k_2\le k_{12} + k_1.
\label{range12}
\end{eqnarray}
Moreover, from the momentum conservation we obtain
$-{\bf k_{12}}={\bf k_3} + {\bf k_4}$, and thus the conditions
\begin{eqnarray}
k_{12}\le k_3 + k_4, \qquad
k_3\le k_{12} + k_4 \qquad \mbox{and} \qquad
k_4\le k_{12} + k_3,
\label{range34}
\end{eqnarray}
must be also imposed.
The integration range of $\cos \theta_4$ is fixed by
inequality~(\ref{const_costheta4}).
Due to the symmetry under interchange of $k_1$ and $k_2$,
we can confine the integration range to $k_1\ge k_2$.
With this choice of weight, the shape correlator is defined as
\begin{eqnarray}
\bar{\mathcal{C}} (S_{\mathcal{T}}, S_{\mathcal{T}'})
= \frac{F( S_{\mathcal{T}}, S_{\mathcal{T}'} )}
{\sqrt{F( S_{\mathcal{T}}, S_{\mathcal{T}} )
F( S_{\mathcal{T}'}, S_{\mathcal{T}'} ) }}\,.
\label{pritrispectrum_correlator}
\end{eqnarray}

\subsection{The shape correlator based on the full trispectrum}
\label{correlator}


In this subsection, we implement the shape correlator
using the full trispectrum $T_\zeta$.
As we mentioned in the previous subsection, while the reduced trispectrum has
all information of the original trispectrum,
the decomposition of the full trispectrum into
the reduced trispectra is not unique.
If all momenta appear symmetrically in the definition of the correlator,
the obtained correlation is independent of the choice of the
reduced trispectrum.
However, the definition given by Eq.~(\ref{pritrispectrum_correlator_part})
apparently breaks the symmetry
because of the integration variables $k_{12}$ and $\theta_4$.
Therefore, in order to define the correlator uniquely,
we must use the full trispectrum in the definition of the shape correlator.

The shape correlator based on the full trispectrum is defined in a similar way with
the one based on the reduced trispectrum.
Namely, we begin with defining the shape function as
\begin{eqnarray}
S_{T}(k_1, k_2, k_3, k_4, k_{12}, \theta_4)
= (k_1 k_2 k_3 k_4)^2 k_{12}
T_\zeta (k_1, k_2, k_3, k_4; k_{12}, \theta_4)\, .
\label{shapefuncT_def}
\end{eqnarray}
The inner product is defined as
\begin{eqnarray}
&&F(S_{T}, S_{T'})\nonumber\\
&&\quad
= \int d k_1 d k_2 d k_3 d k_4 d k_{12} \int d (\cos \theta_4)
S_{T} (k_1, k_2, k_3, k_4, k_{12}, \theta_4)
S_{T'} (k_1, k_2, k_3, k_4, k_{12}, \theta_4)
w(k_1, k_2, k_3, k_4, k_{12})\, .
\label{pritrispectrumT_correlator_part}
\end{eqnarray}
where we use the same weight function $w$ as that in the case of reduced trispectrum.
The domain of integration in Eq.~(\ref{pritrispectrumT_correlator_part}) is basically
the same as that for the reduced trispectrum, i.e.
inequalities~(\ref{const_costheta4}), (\ref{range12}) and (\ref{range34}).
However, since the symmetry under interchange of $k_1$ and $k_2$ still remains,
the additional condition $k_1\ge k_2$ does not change the final value of the shape correlator.Therefore, we adopt inequalities~(\ref{const_costheta4}),  (\ref{range12}),
(\ref{range34}) and $k_1\ge k_2$ as the domain of integration.

The shape correlator is defined in the same way
\begin{eqnarray}
\bar{\mathcal{C}} (S_T, S_{T'})
= \frac{F( S_{T}, S_{T'} )}
{\sqrt{F( S_{T}, S_{T} )
F( S_{T'}, S_{T'} ) }}\,.
\label{pritrispectrumT_correlator}
\end{eqnarray}

\section{Shape correlations}
\label{Shapecorrelations}

In this section,
we explicitly calculate the shape correlations among
trispectra predicted by various theoretical models.
The main purpose of investing the trispectrum
is discriminating between models which are hard to be 
distinguished by the bispectrum. Thus we concentrate on
the shape correlations among the models where the bispectrum is
dominated by the equilateral type one; single field and multi-field
DBI inflation models, ghost inflation model and
Lifshitz scalar model. 

The explicit forms of the shape functions are calculated in Appendixes. 
$ S_\mathcal{T}^{DBI(\sigma)} $, $ S_\mathcal{T}^{DBI(s)} $ and 
$ S_\mathcal{T}^{ghost} $ are defined as the shape functions in single field  
DBI Inflation, multi-field DBI inflation and ghost inflation, respectively. 
Explicit forms of them are written in Eqs.(\ref{singleshapefunction}), 
(\ref{multishapefunction}) and (\ref{ghostshapefunction}). In Lifshitz scalar field, 
there are a few contributions of Trispectrum and it depends on the 
parameters in the theory. We define $S_\mathcal{T}^{h(se,11)}$, 
$S_\mathcal{T}^{h(se,12)}$, $S_\mathcal{T}^{h(se,13)}$, $S_\mathcal{T}^{h(ci,1)}$, 
$S_\mathcal{T}^{h(ci,2)}$, and $S_\mathcal{T}^{h(ci,3)}$ as the shape functions of 
respective contributions. We show explicit forms of them in 
Eqs.(\ref{horavaseshapefunction}) and (\ref{Sci}).

Furthermore, the shape given by $S^{c1}_{\mathcal{T}}$
(see Eq.~(\ref{shape_c1})) can be written in a separable form
\cite{Chen:2009bc} and it provides a fast estimator for the
equilateral type trispectrum
\cite{2010arXiv1004.2915R,Mizuno:2010by}.
This means that if the shape function for the trispectrum
is highly correlated with $S^{c1}_{\mathcal{T}}$,
we can constrain the model parameters from the amplitude
of the trispectrum in a very simple and fast way
as we will discuss in Sec.~IV. 
Therefore, we also calculate the shape
correlations of the trispectra predicted by theories
mentioned above with $S^{c1}_{\mathcal{T}}$.

Another motivation for calculating the shape
correlations is to check the difference
between the two shape correlators that we introduced
in the previous section.
As we mentioned in the previous section, the shape correlation based
on the reduced trispectrum  depends on the way to
decompose the full trispectrum into the reduced trispectra.
Thus strictly speaking, we should use the full tripsectrum
when calculating the shape correlation.
However, the calculation is easier for the shape correlation
using the reduced trispectrum.
If the difference is insignificant, we may still use the
reduced trispectrum to calculate the shape correlator.

\subsection{Shape correlations based on the reduced trispectrum}

\begin{figure}[t]
  \begin{center}
    \includegraphics[keepaspectratio=true,height=20mm]{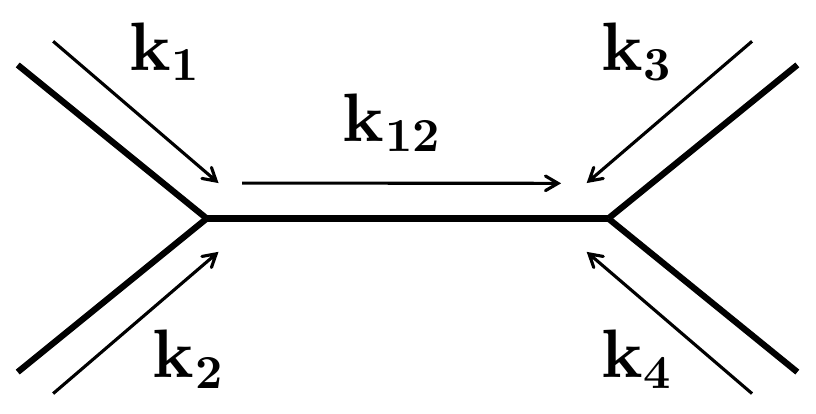}
  \end{center}
  \caption{scalar exchange diagram with $k_{12}$ inner propagator}
  \label{fig:se.eps}
\end{figure}

Here, we calculate the shape correlations among
the theoretical models based on the reduced trispectrum.
We must specify how we decompose the full trispectrum
into reduced trispectra.
One natural way is to decompose it so that
one of the reduced trispectrum depends
only on five parameters $k_1$, $k_2$, $k_3$, $k_4$ and
$k_{12}$.
Except for the case of the trispectrum
obtained by the contact interaction in Lifshitz scalar model
for $i=3$ in Eq.~(\ref{Sci}),
it is possible to find such a decomposition. In fact,
in the case of the scalar exchange trispectrum, this is always the case.
The contribution of the trispectrum from the scalar exchange can be decomposed into
s-, t- and u-channels and one of the channel depends only on five parameters
$k_1$, $k_2$, $k_3$, $k_4$ and $k_{12}$. This can be understood as follows.
The diagram of the scalar exchange trispectrum has one inner propagator.
In the diagram where the momentum of the inner propagator is $k_{12}$,
the propagators with momenta $k_1$ and $k_2$ meet directly
at one vertex. On the other hand, the propagators with momenta $k_3$ and $k_4$ 
meet at the other vertex (see Fig.~\ref{fig:se.eps}).
Then, the trispectrum from Fig.~\ref{fig:se.eps} is written
in terms of  $k_1$, $k_2$, $k_3$, $k_4$,
$k_{12}$, $\bf k_1 \cdot k_2$ and $\bf k_3 \cdot k_4$.
The inner products $\bf k_1 \cdot k_2$ and $\bf k_3 \cdot k_4$ can be expressed as
$(k_{12}^2-k_1^2-k_2^2)/2$ and $(k_{12}^2-k_3^2-k_4^2)/2$, respectively.
Therefore, the trispectrum from Fig.~\ref{fig:se.eps} depends only on
$k_1$, $k_2$, $k_3$, $k_4$ and $k_{12}$.
Moreover, in most models even the contact interaction trispectrum can be also decomposed
in such a way that the reduced trispectrum only depends on $k_1$, $k_2$, $k_3$, $k_4$ and $k_{12}$. In the contact interaction, there are no inner propagators. Thus
the dependence on $k_{12}$, $k_{13}$ and
$k_{14}$ stem only from the derivative coupling such as
$(\vec \nabla_i \phi)^2 \phi^2$.
Only when the product of $k_{12}$ and $k_{13}$
appear in a contact interaction trispectrum,
the trispectrum can not be decomposed in the way so that reduced trispectrum
depends only on $k_1$, $k_2$, $k_3$, $k_4$ and $k_{12}$.
This actually happens for the contact interaction
in Lifshitz scalar field for $i=3$ in Eq.~(\ref{Sci}).

In appendix A, B, C, we summarise the explicit expressions of
the shape functions defined by the reduced trispectra
in single field and multi-field DBI inflation models,
ghost inflation model and Lifshitz scalar model,
respectively. As is mentioned above, except
for the one corresponding to the contact interaction
in Lifshitz scalar field for $i=3$ in (\ref{Sci}),
they depend only on five parameters $k_1$, $k_2$,
$k_3$, $k_4$ and $k_{12}$.
For the trispectrum coming from the contact interaction in Lifshitz scalar
field for $i=3$ in Eq.~(\ref{Sci}), we need to define the reduced trispectrum
so that it is symmetric under the transpose among $\boldsymbol{k_1}$, $\boldsymbol{k_2}$,
$\boldsymbol{k_3}$ and $\boldsymbol{k_4}$.

The shape correlations among the models are
summarised in Table~\ref{Table;reduced}.
From this table, $S^{DBI(\sigma)}_{\mathcal{T}}$
and $S^{h(se,22)}_{\mathcal{T}} $ turn out to
be highly correlated with  $S^{c1}_{\mathcal{T}}$.

\subsection{Shape correlations based on the full trispectrum}

\begin{figure}[t]
  \begin{center}
    \includegraphics[keepaspectratio=true,height=50mm]{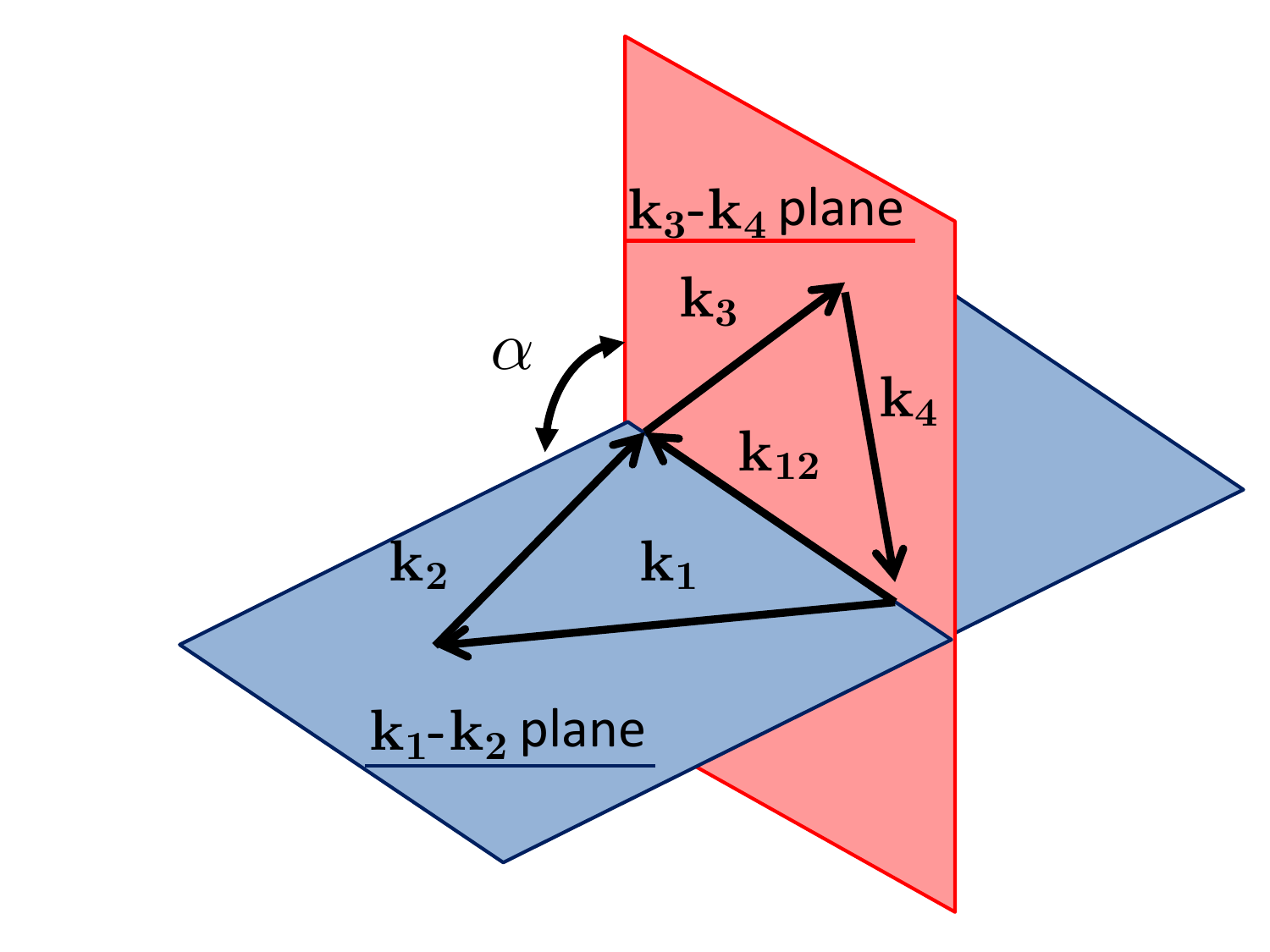}
  \end{center}
  \caption{definition of $\alpha$}
  \label{fig:alpha.eps}
\end{figure}

Now, we move to the shape correlations based on the
full trispectrum.
With Eqs.~(\ref{deftri}) and (\ref{defrtri}),
the full trispectrum is obtained from
the reduced trispectrum summarised in the appendix sections,
by adding their permutations $(k_2 \leftrightarrow k_3)$
and $(k_2 \leftrightarrow k_4)$.
Then, the shape functions based on the full trispectrum
can be written by $k_1$, $k_2$,
$k_3$, $k_4$, $k_{12}$, $k_{13}$ and $k_{14}$.
We substitute the shape functions of the full trispectra
into the definition of the shape correlator (\ref{pritrispectrumT_correlator}).
Thus, we need to express $k_{13}$ and $k_{14}$
in terms of $k_1$, $k_2$,
$k_3$, $k_4$, $k_{12}$ and $\theta_4$.
We find that these variables are expressed as
\begin{eqnarray}
k_{13} &=&\biggl\{( k_2^2 + k_4^2 - \frac{1}{2 k_{12}^2}
(k_2^2 + k_{12}^2 - k_1^2)(k_4^2 + k_{12}^2 - k_3^2)
\nonumber\\
&&\mp \frac{1}{2 k_{12}^2}
\sqrt{4 k_2^2 k_{12}^2 - (k_2^2 + k_{12}^2 - k_1^2)^2}
\sqrt{4 k_4^2 k_{12}^2 \cos^2 \theta_4 -
(k_4^2 + k_{12}^2-k_3^2)^2})\biggr\}^{1/2} \,,\nonumber\\
k_{14} &=&\biggl\{( k_1^2 + k_4^2 - \frac{1}{2 k_{12}^2}
(k_1^2 + k_{12}^2 - k_2^2)(k_4^2 + k_{12}^2 - k_3^2)
\nonumber\\
&&\mp \frac{1}{2 k_{12}^2}
\sqrt{4 k_1^2 k_{12}^2 - (k_1^2 + k_{12}^2 - k_2^2)^2}
\sqrt{4 k_4^2 k_{12}^2 \cos^2 \theta_4 -
(k_4^2 + k_{12}^2-k_3^2)^2})\biggr\}^{1/2} \, ,
\label{k14_ito_123412costheta4}
\end{eqnarray}
where $\pm$ is determined by the configurations of the momenta;
if the angle between the planes which are specified by $(k_1,k_2,k_{12})$-triangle and
$(k_3,k_4,k_{12})$-triangle, $\alpha$, is smaller than the right angle
(see Fig.~\ref{fig:alpha.eps}),
the corresponding sign is the minus ($-$).
Otherwise, we should choose the plus ($+$).

\begin{table}[t]
 \caption{The shape correlations among the models
based on the reduced trispectra}
 \begin{center}
  \begin{tabular}{|c|c|c|c|c|c|c|c|c|c|}
    \hline
       \,  & $S^{DBI(\sigma)}_{\mathcal{T}}$   &
    $ S^{DBI(s)}_{\mathcal{T}}   $ & $ S^{ghost}_{\mathcal{T}}  $&
    $ S^{h(se,11)}_{\mathcal{T}} $ & $S^{h(se,12)}_{\mathcal{T}} $  &
    $ S^{h(se,22)}_{\mathcal{T}} $  & $S^{h(ci,1)}_{\mathcal{T}}$   &
    $ S^{h(ci,2)}_{\mathcal{T}}  $  &$ S^{h(ci,3)}_{\mathcal{T}}  $   \\
    \hline
   $   S^{c1}_{\mathcal{T}}$ & 0.87 &0.33 & 0.24   &  0.24  &
      -0.62   &  0.95  &  0.35  &  0.32  &0.53    \\
    \hline
   $   S^{DBI(\sigma)}_{\mathcal{T}}$ & &0.19 &  0.41 & 0.42   &  -0.78  &
      0.96   &  0.60  &  0.48   &0.68     \\
    \hline
    $  S^{DBI(s)}_{\mathcal{T}} $&  &  &  0.59  &0.13  & -0.25 &0.28 &0.21 &  0.42 &-0.78\\
    \hline
    $ S^{ghost}_{\mathcal{T}} $ &  &  &  & 0.23 & -0.31 & 0.30 & 0.86 & 0.90&-0.36 \\
    \hline
    $ S^{h(se,11)}_{\mathcal{T}} $  &  &  &  &  &-0.75 & 0.33 & 0.37 & 0.15 &0.25   \\
    \hline
     $ S^{h(se,12)}_{\mathcal{T}} $&  &  &  &  &  & -0.73 & -0.50 &  -0.32&-0.52 \\
    \hline
   $  S^{h(se,22)}_{\mathcal{T}} $ &  &  &  &  &  &  & 0.44 &  0.39 &0.63 \\
    \hline
    $  S^{h(ci,1)}_{\mathcal{T}}  $&  &  &  &  &  &  &  & 0.88&0.19 \\
    \hline
    $  S^{h(ci,2)}_{\mathcal{T}}  $&  &  &  &  &  &  &  & &-0.01 \\
    \hline
  \end{tabular}
  \label{Table;reduced}
 \end{center}
\end{table}

\begin{table}[t]
 \caption{The shape correlations among the models based
on the full trispectra}
 \begin{center}
  \begin{tabular}{|c|c|c|c|c|c|c|c|c|c|}
    \hline
       \,  & $S^{DBI(\sigma)}_T$   &
    $ S^{DBI(s)}_T   $ & $ S^{ghost}_T  $&
    $ S^{h(se,11)}_T $ & $S^{h(se,12)}_T $  &
    $ S^{h(se,22)}_T $  & $S^{h(ci,1)}_T$   &
    $ S^{h(ci,2)}_T  $ &$ S^{h(ci,3)}_T  $   \\
    \hline
   $   S^{c1}_T$ &0.96 &0.42&0.41&0.24&-0.70&0.99&0.54&0.52&0.38   \\
    \hline
   $   S^{DBI(\sigma)}_T$ & &0.40&0.51&0.36&-0.80&0.98&0.71&0.62&0.43    \\
    \hline
    $  S^{DBI(s)}_T$&  &  & 0.76&0.25&-0.45&0.45&0.39&0.60&-0.40 \\
    \hline
    $ S^{ghost}_T$ &  &  &  & 0.24&-0.42&0.48&0.83&0.92&-0.10 \\
    \hline
    $ S^{h(se,11)}_T $  &  &  &  &  & - 0.70&0.32&0.36&0.19&0.06 \\
    \hline
     $ S^{h(se,12)}_T $&  &  &  &  &  &-0.77&-0.60&-0.45& -0.22 \\
    \hline
   $  S^{h(se,22)}_T$ &  &  &  &  &  &  & 0.62&0.58&0.38\\
    \hline
    $  S^{h(ci,1)}_T  $&  &  &  &  &  &  &  & 0.90&0.28\\
    \hline
    $  S^{h(ci,2)}_T  $&  &  &  &  &  &  &  &  &0.09\\
    \hline
  \end{tabular}
  \label{Table;full}
 \end{center}
\end{table}

The shape correlations among the models are
summarised in Table~\ref{Table;full}.
In most of the models, the shape correlations based
on the reduced trispectra are
close to those based on the full trispectra.
Especially, if the correlation based on the reduced trispectra
is high, there is a little difference between the two shape correlations. 
Thus as long as the shape correlation is high, it is possible to use the reduced
trispectrum to calculate the shape correlation. On the other hand,
a care must be taken if the correlation is low.


\section{Theoretical predictions and observational constraints}

In this section, in order to constrain the model parameters
from the amplitude of the trispectrum,
we will calculate the amplitude of the estimators 
$g_{\rm NL} ^{equil}$ based on the shape $S_{\cal T}^{c1}$ (see Eq.~(\ref{shape_c1})),
which can be written in a separable form \cite{Chen:2009bc}.
We also discuss the amplitude of the trispectrum in the regular tetrahedron limit,
$t_{\rm NL}$ and the maximally symmetric configuration,
$\tau_{\rm NL}$.
  
\subsection{$g_{\rm NL}^{equil}$}
\label{gNL}

Here, following Ref.~\cite{Mizuno:2010by}, we discuss
the theoretical predictions
for the amplitude of the trispectrum $g_{\rm NL}^{equil}$.
We first define the equilateral shape using $S_{\cal T}^{c1}$,
\begin{eqnarray}
S_{\cal T}^{equil}= \frac{64}{3} (2\pi^2 {\cal P}_\zeta)^3 S_{\cal T}^{c1},
\end{eqnarray}
where ${\cal P}_\zeta$ is defined as
$\left\langle \zeta (\k_1) \zeta_(\k_2) \right\rangle=
16 \pi^5  \delta^3(\k_1+\k_2) k_1^{-3}{\cal P}_\zeta$.
The parameter $g_{\rm NL}^{equil}$ is defined as the normalised shape correlation between $S_{\cal T}$ and $S_{\cal T}^{equil}$;
 \begin{eqnarray}
g_{\rm NL}^{equil} \equiv \frac{F(S_{\cal T},S_{\cal T}^{equil})}
{F(S_{\cal T}^{equil},S_{\cal T}^{equil})} .
\label{defg}
\end{eqnarray}

We explicitly calculate $g_{\rm NL}^{equil}$ in single field DBI inflation model,
multi-field DBI inflation model, ghost inflation model and
Lifshitz scalar model.
The results are summarised in Table~\ref{Table;gnl}.
Since we have two kinds of the shape correlators
(one is based on the reduced trispectrum
and the other is based on the full trispectrum),
we show the two results in Table~\ref{Table;gnl}.
Here, we use the fact that the power specrta of the primordial perturbations
in single field and multi-field DBI inflation models,
ghost inflation model and Lifshitz scalar model
are given by
\begin{eqnarray}
&&\mathcal{P}_\zeta^{\sigma}=\frac{1}{2\pi^2} \frac{H^4}{2 \dot\phi^2}, \\
&&\mathcal{P}_\zeta^{s}=\frac{1}{2\pi^2} \frac{H^4 T_{\cal RS}}{2 \dot\phi^2},\\
&&\mathcal{P}_\zeta^{ghost}=\frac{1}{\pi \left(\Gamma(1/4)\right)^2 \alpha^{3/4}}
\left(\frac{H}{M}\right)^{\frac{5}{2}},\\
&&\mathcal{P}_\zeta^{h}=\frac{1}{2\pi^2}\frac{M^2}{2\mu^2}.
\end{eqnarray}
Moreover, in Lifshitz scalar model, we individually show the contributions
from each shape function.

Notice that although we obtain $g_{\rm NL} ^{equil}$ for
all theoretical models mentioned above, this $g_{\rm NL} ^{equil}$ can be
meaningful only in the case where the shape correlation with the equilateral shape
is sufficiently high, like $> 0.7$. Otherwise, this template based on
the equilateral shape $S_{\cal T}^{equil}$ does not fit the shape of
the trispectrum well and the noise dominates over the signal
\footnote{We thank D.~Regan for pointing out this point.}.

From this table, we find that there are little differences between
$g_{\rm NL}^{equil}$ based on
the reduced trispectrum and that based on the full trispectrum
in most of shape functions, especially for $S_{\cal T}^{DBI (\sigma)}$,
$S_{\cal T}^{h(se,12)}$ and $S_{\cal T}^{h(se,22)}$ where
the shape correlation with $S_{\cal T}^{equil}$ is high enough
and the use of $g_{\rm NL} ^{equil}$ to measure the amplitude is justified.


\begin{table}[t]
 \caption{The value of $g_{\rm NL}^{equil}$ based on reduced trispectrum and
 full trispectrum.}
 \begin{center}
  \begin{tabular}{|c|c|c|c|c|c|}
    \hline
       \,  & $S^{DBI(\sigma)}_T$   &
    $ S^{DBI(s)}_T   $ & $ S^{ghost}_T  $  &    $ S^{h(se,11)}_T $ & $S^{h(se,12)}_T $   \\
    \hline
  reduced &$1.4\times10\frac{1}{c_s^4}$&$2.3\frac{1}{c_s^4 
T_{\cal RS}^2}$&
  $4.9\times10^5 \frac{\tilde \gamma}{\alpha^{8/5}}\left(
\frac{\mathcal{P}_\zeta^{1/2}}{4.8\times10^{-5}}\right)^{-8/5}$
  &$1.6\times10^7b_1^2 \left( \frac{\mathcal{P}_\zeta^{1/2}}{4.8\times10^{-5}}\right)^{-2}$&
  $-1.8\times10^8b_1 b_2 \left( \frac{\mathcal{P}_\zeta^{1/2}}{4.8\times10^{-5}}\right)^{-2}$ \\
    \hline
   full &$1.3\times10\frac{1}{c_s^4}$&$2.7\frac{1}{c_s^4
T_{\cal RS}^2}$&
   $6.0\times10^5\frac{\tilde \gamma}{\alpha^{8/5}}\left(
\frac{\mathcal{P}_\zeta^{1/2}}{4.8\times10^{-5}}\right)^{-8/5}$ &$1.1
       \times10^7b_1^2 \left(
       \frac{\mathcal{P}_\zeta^{1/2}}{4.8\times10^{-5}}\right)^{-2}$&$-1.6\times10^8b_1
       b_2 \left(
\frac{\mathcal{P}_\zeta^{1/2}}{4.8\times10^{-5}}\right)^{-2}$ \\
    \hline
  \end{tabular}
 \end{center}
  \begin{center}
  \begin{tabular}{|c|c|c|c|c|}
    \hline
       \,  &
    $ S^{h(se,22)}_T $  & $S^{h(ci,1)}_T$   &
    $ S^{h(ci,2)}_T  $ &$ S^{h(ci,3)}_T  $   \\
    \hline
  reduced &$4.5\times10^8b_2^2 \left(
       \frac{\mathcal{P}_\zeta^{1/2}}{4.8\times10^{-5}}\right)^{-2}$&$9.1\times10^7t_1
       \left(
       \frac{\mathcal{P}_\zeta^{1/2}}{4.8\times10^{-5}}\right)^{-2}$&$2.0\times10^8t_2
       \left(
       \frac{\mathcal{P}_\zeta^{1/2}}{4.8\times10^{-5}}\right)^{-2}$&$3.7\times10^8t_3
       \left(
\frac{\mathcal{P}_\zeta^{1/2}}{4.8\times10^{-5}}\right)^{-2}$   \\
    \hline
   full&$4.3\times10^8b_2^2 \left( \frac{\mathcal{P}_\zeta^{1/2}}{4.8\times10^{-5}}\right)^{-2}$&$9.3\times10^7t_1 \left( \frac{\mathcal{P}_\zeta^{1/2}}{4.8\times10^{-5}}\right)^{-2}$&$2.2\times10^8t_2 \left( \frac{\mathcal{P}_\zeta^{1/2}}{4.8\times10^{-5}}\right)^{-2}$&$6.2\times10^7t_3 \left( \frac{\mathcal{P}_\zeta^{1/2}}{4.8\times10^{-5}}\right)^{-2}$   \\
    \hline
  \end{tabular}
  \label{Table;gnl}
 \end{center}
\end{table}

\subsection{$t_{\rm NL}$ and $\tau_{NL}$}
\label{tNL}

In order to estimate the amplitude of the trispectrum with 
different shapes, Chen et al. \cite{Chen:2009bc} define
the amplitude of the trispectrum $t_{\rm NL}$ using
a particular configuration as follows;
\begin{eqnarray}
\langle \zeta^4 \rangle \to (2 \pi)^9 \mathcal{P}_\zeta ^3 \delta^3(\sum_i
{\bold k_i}) \frac{1}{k^9} t_{\rm NL},
\label{def_tnl}
\end{eqnarray}
where the limit stands for the regular tetrahedron limit
$(k_1=k_2=k_3=k_4=k_{12}=k_{13} =k)$.
Theoretical predictions for $t_{\rm NL}$ are
summarised in Tables~\ref{Table;tnl}.

It is clear that $t_{\rm NL}$ gives only a rough estimation of
the amplitude of the trispectrum as it is obtained by using
only one specific configuration. In fact, for $S^{h(ci,3)}_T$,
$t_{\rm NL}$ is $0$ because  $S^{h(ci,3)}_T$ happens to
vanish for the configuration given by $(k_1=k_2=k_3=k_4=k_{12}=k_{13} =k)$.
On the other hand, if we consider all the configurations, there is
a small but non-negligible over-lap between this shape and
$S_{\cal T}^{c1}$. This clearly demonstrates that it is necessary to use
all the configurations to estimate the amplitude of the trispectrum.

We also show the amplitude of the trispectrum in another definition. 
The configuration $(k_1=k_2=k_3=k_4 =k$ and 
$k_{12}=k_{13}=k_{14}=2 k /\sqrt{3})$ is maximally symmetric 
while the configuration of tetrahedron is not invariant
under interchange of $\bf k_1$ and $\bf k_2$.
Therefore, the maximally symmetric configuration is probably useful to 
analyze the trispectrum. 
Here, we define the amplitude of the trispectrum $\tau_{\rm NL}$ as in \cite{Arroja:2009pd};
\footnote{While in \cite{Arroja:2009pd} the amplitude of the trispectrum $\tau_{\rm NL}$ 
is defined in any configuration, 
we rewrite the simplified form which can be applied only in the maximally symmetric case.}
\begin{eqnarray}
\tau_{NL}=\frac{2}{9\sqrt{3}} k^9 \frac{T_\zeta}{(2 \pi^2 {\mathcal{P}_\zeta})^3},
\end{eqnarray}
where $T_\zeta$ is the trispectrum of the maximally symmetric configuration. 
Theoretial predictions for $\tau_{NL}$  are also
summarised in Tables~\ref{Table;tnl}.

\begin{table}[t]
 \caption{The value of $t_{\rm NL}$ and $\tau_{NL}$.}
 \begin{center}
  \begin{tabular}{|c|c|c|c|c|c|}
    \hline
       \,  & $S^{DBI(\sigma)}_T$   &
    $ S^{DBI(s)}_T   $ & $ S^{ghost}_T  $  &    $ S^{h(se,11)}_T $ & $S^{h(se,12)}_T $   \\
    \hline
  $t_{\rm NL}$ &$5.4\times10^{-1}\frac{1}{c_s^4}$
  &$1.4 \times 10^{-1}\frac{1}{c_s^4T_{\cal RS}^2}$
  & $-1.2 \times 10^{5} \frac{\tilde \gamma}{\alpha^{8/5}}\left( \frac{\mathcal{P}_\zeta^{1/2}}{4.8\times10^{-5}}\right)^{-8/5}$ &  $5.1\times 10^6 b_1^2 \left( \frac{\mathcal{P}_\zeta^{1/2}}{4.8\times10^{-5}}\right)^{-2}$& $-2.3\times10^7 b_1 b_2 \left( \frac{\mathcal{P}_\zeta^{1/2}}{4.8\times10^{-5}}\right)^{-2}$\\
    \hline  
    $\tau_{\rm NL}$ &$5.6\times 10^{-1} \frac{1}{c_s ^4}$
  &$1.2 \times 10^{-1}\frac{1}{c_s^4T_{\cal RS}^2}$
  & $-1.0 \times 10^{4} \frac{\tilde \gamma}{\alpha^{8/5}}\left( \frac{\mathcal{P}_\zeta^{1/2}}{4.8\times10^{-5}}\right)^{-8/5}$ &  $1.2\times 10^7 b_1^2 \left( \frac{\mathcal{P}_\zeta^{1/2}}{4.8\times10^{-5}}\right)^{-2}$& $-9.9\times10^7 b_1 b_2 \left( \frac{\mathcal{P}_\zeta^{1/2}}{4.8\times10^{-5}}\right)^{-2}$\\
    \hline
  \end{tabular}
 \end{center}
  \begin{center}
  \begin{tabular}{|c|c|c|c|c|}
    \hline
       \,  &
    $ S^{h(se,22)}_T $  & $S^{h(ci,1)}_T$   &
    $ S^{h(ci,2)}_T  $ &$ S^{h(ci,3)}_T  $   \\
    \hline
  $\tau_{\rm NL}$ &  $2.8\times 10^7 b_2^2 \left(
\frac{\mathcal{P}_\zeta^{1/2}}{4.8\times10^{-5}}\right)^{-2}$&
       $2.7\times 10 ^6 t_1 \left(
       \frac{\mathcal{P}_\zeta^{1/2}}{4.8\times10^{-5}}\right)^{-2}$&$
       6.9\times 10^5 t_2 \left(
\frac{\mathcal{P}_\zeta^{1/2}}{4.8\times10^{-5}}\right)^{-2}$& 0   \\
    \hline
      $t_{\rm NL}$ &  $2.1\times 10^8 b_2^2 \left(
\frac{\mathcal{P}_\zeta^{1/2}}{4.8\times10^{-5}}\right)^{-2}$&
       $2.8\times 10 ^6 t_1 \left(
       \frac{\mathcal{P}_\zeta^{1/2}}{4.8\times10^{-5}}\right)^{-2}$&
       $ 3.1\times 10^5 t_2 \left(
\frac{\mathcal{P}_\zeta^{1/2}}{4.8\times10^{-5}}\right)^{-2}$& 
     $3.1\times 10 ^5 t_3 \left(
       \frac{\mathcal{P}_\zeta^{1/2}}{4.8\times10^{-5}}\right)^{-2}$   \\
    \hline
  \end{tabular}
  \label{Table;tnl}
 \end{center}
\end{table}

\subsection{Constraints on model parameters}
Observational constraints on the equilateral non-Gaussianity
were obtained from WMAP5 data by applying the estimator with the shape given by $S_{\cal T}^{c1}$ \cite{Fergusson:2010gn}.
The constraints on the amplitude of the trispectrum in the regular tetrahedron
limit is obtained as  
\begin{eqnarray}
t_{\rm NL} ^{equil}= (-3.11 \pm 7.5) \times 10^6\;(68\% {\rm CL}).
\label{const_tnl}
\end{eqnarray}
This constraint can be converted to that on $g_{\rm NL}^{equil}$ as follows.
We define the following shape function
\begin{equation}
S_{{\cal T}} =g_{\rm NL}^{equil} \frac{64}{3} (2\pi^2 {\cal P}_\zeta)^3 S_{\cal T}^{c1},
\end{equation}
which by definition gives $g_{\rm NL}^{equil}$ when we apply Eq.~(\ref{defg}). Now using the relation between the shape function and the full trispectrum and evaluating the full trispectrum in the regular tetrahedron limit, we can calculate $t_{\rm NL}^{equil}$ for this trispectrum as 
\begin{equation}
t_{\rm NL} ^{equil} = \frac{1}{32} g_{\rm NL}^{equil}.
\end{equation}
Then we obtain the constraints on $g_{\rm NL}^{equil}$ as
\begin{eqnarray}
g_{\rm NL} ^{equil}= (-9.95\pm 24) \times 10^7\;(68\% {\rm CL}).
\label{const_gnl}
\end{eqnarray}
From the theoretical predictions for $g_{\rm NL} ^{equil}$ in various
models obtained by using the full trispectrum
as well as the reduced trispectrum, we can derive the constraints on model parameters as in Table~\ref{Table;constraint}.

\begin{table}[t]
 \caption{Constraints on model parameters from
$g_{\rm NL}^{equil}$ in various models.}
 \begin{center}
  \begin{tabular}{|c|c|c|}
  \hline
   & reduced & full \\
  \hline
       $S^{DBI(\sigma)}_T$   &$1.8 \times 10^{-2} < c_s$ &
     $1.7 \times 10^{-2} < c_s$ \\
  \hline  
    $ S^{DBI(s)}_T   $ & $1.1 \times 10^{-2} < c_s \sqrt{T_{\cal RS}}$&
    $1.2 \times 10^{-2} < c_s \sqrt{T_{\cal RS}}$ \\
  \hline  
    $ S^{ghost}_T  $  &    $-6.9 \times 10^{2} <  \frac{\tilde \gamma}{\alpha^{8/5}}
  \left( \frac{\mathcal{P}_\zeta^{1/2}}{4.8\times10^{-5}}\right)^{-8/5}
  < 2.9 \times 10^{2} $
  & 
   $-5.7 \times 10^{2} <  \frac{\tilde \gamma}{\alpha^{8/5}}
  \left( \frac{\mathcal{P}_\zeta^{1/2}}{4.8\times10^{-5}}\right)^{-8/5}
  < 2.3 \times 10^{2} $ 
  \\ 
  \hline
  $ S^{h(se,11)}_T $  &
  $(0<)\;b_1^2 \left(    
   \frac{\mathcal{P}_\zeta^{1/2}}{4.8\times10^{-5}}\right)^{-2} < 8.8 $
   &
   $(0<)\;b_1^2 \left(    
   \frac{\mathcal{P}_\zeta^{1/2}}{4.8\times10^{-5}}\right)^{-2} < 1.3 \times10$
   \\
    \hline
        $S^{h(se,12)}_T $ & $-7.8 \times 10^{-1} < b_1 b_2 \left(
    \frac{\mathcal{P}_\zeta^{1/2}}{4.8\times10^{-5}}\right)^{-2} < 1.9$
   &
   $-8.8 \times10^{-1} < b_1 b_2 \left(
    \frac{\mathcal{P}_\zeta^{1/2}}{4.8\times10^{-5}}\right)^{-2} < 2.1$
   \\
    \hline
    $ S^{h(se,22)}_T $  & $(0<)\;b_2^2 \left(
    \frac{\mathcal{P}_\zeta^{1/2}}{4.8\times10^{-5}}\right)^{-2}<3.1 \times10^{-1}$
    &$(0<)\;b_2^2 \left(
    \frac{\mathcal{P}_\zeta^{1/2}}{4.8\times10^{-5}}\right)^{-2}<3.3 \times10^{-1}$ 
    \\    
    \hline
    $S^{h(ci,1)}_T$&  
    $-3.7
    < t_1 \left( \frac{\mathcal{P}_\zeta^{1/2}}{4.8\times10^{-5}}\right)^{-2}
    < 1.5 $
    &     
    $-3.7 
    < t_1 \left( \frac{\mathcal{P}_\zeta^{1/2}}{4.8\times10^{-5}}\right)^{-2}
    < 1.5 $  \\
    \hline
    $ S^{h(ci,2)}_T  $ &$-1.7  < t_2 \left(
       \frac{\mathcal{P}_\zeta^{1/2}}{4.8\times10^{-5}}\right)^{-2} < 7.0\times 10^{-1}$
       &$-1.5 < t_2 \left(
       \frac{\mathcal{P}_\zeta^{1/2}}{4.8\times10^{-5}}\right)^{-2} < 6.4\times10^{-1}$
       \\
    \hline
    $ S^{h(ci,3)}_T  $ &  
      $-9.2 \times 10^{-1} < t_3 \left(
       \frac{\mathcal{P}_\zeta^{1/2}}{4.8\times10^{-5}}\right)^{-2} < 
       3.8 \times 10^{-1}$
       & $-5.5 < t_3 \left(
       \frac{\mathcal{P}_\zeta^{1/2}}{4.8\times10^{-5}}\right)^{-2} < 
       2.3$
         \\
    \hline
  \end{tabular}
  \label{Table;constraint}
 \end{center}
\end{table}

\section{Summary and Discussions}
\label{sec:summary}

There are many interesting early universe models motivated by string theory 
and the effective field theory that predict large primordial
non-Gaussianity with the equilateral type bispectrum.
Given that future experiments like Planck can prove even higher-order
statistics, it is important to investigate whether
the shape of the trispectrum can distinguish
such equilateral type non-Gaussian models.
For this purpose, the shape correlator of the trispectrum
is particularly useful.

So far, the shape correlator constructed from the reduced trispectrum has been 
adopted \cite{2010arXiv1004.2915R}.
While the reduced trispectrum has all information of the full trispectrum,
the shape correlator based on the reduced trispectrum depends on the way
in which the full trispectrum is decomposed into reduced trispectra
because the form of the integration variables in the shape correlator breaks
the transposition invariance of momenta $\bf k_1$, $\bf k_2$, $\bf k_3$ and $\bf k_4$.
In most equilateral type non-Gaussian models
such as DBI inflation model, there is a natural way to decompose
the full trispectrum into the reduced trispectra so that one of the reduced
trispectra depends only on five parameters $k_1$, $k_2$,
$k_3$, $k_4$ and $k_{12}$.
However, for some classes of equilateral type non-Gaussian
model like Lifshitz scalar model,
this decomposition is not possible.
Therefore, in this paper, we studied the shape correlator of the primordial trispectrum
based on the full trispectrum.


In order to check the difference between the two 
shape correlators, we calculated the shape correlations
among trispectra in various equilateral non-Gaussian models;
DBI inflation, ghost inflation and Lifshitz scalar models.
We found that both shape correlators give similar results
as long as the shape correlations are high.

From the shape correlations, it is possible to judge whether
we can distinguish between various equilateral non-Gaussian models
using the tripsectrum. For example, we showed that
it is difficult to distinguish between
the single field DBI inflation model and the 
$b_2b_2$-dominated Lifshitz scalar model
by the shape of the trispectrum.
Since both single field DBI inflation model
\cite{Alishahiha:2004eh} and
$b_2b_2$-dominated Lifshitz scalar model \cite{Izumi:2010yn}
predict almost the same equilateral type bispectrum,
we can not distinguish between these two models by
the primordial non-Gaussianity up to this order.
In order to distinguish between these models,
we need other information such as
the primordial gravitational wave.
A similar conclusion holds for the comparison between
the ghost inflation model and $t_1$-dominated
Lifshitz scalar model. Despite these exceptions,
our result suggests that we can distinguish
between many equilateral non-Gaussian models, which predict almost the same equilateral type bispectrum, from the shape of the trispectrum.

On the other hand, in order to measure the amplitude
of the trispectrum and constrain parameters in the theoretical
models, it is necessary to develop an estimator for
the trispectrum. Since the form of the trispectrum
is too complicated for this class of models,
it is generally impossible to construct an optimal estimator (see however \cite{2010arXiv1004.2915R} for the model independent approach).
Using the fact that the shape $S_{\mathcal{T}} ^{c1}$ given by
Eq.~(\ref{shape_c1}) can be written as a separable form,
it is possible to construct a fast optimal estimator
$g_{\rm NL} ^{equil}$  \cite{Mizuno:2010by,Fergusson:2010gn}.
We expressed the estimator in terms of models parameters in various 
models. The constraint on $g_{\rm NL} ^{equil}$ was obtained from WMAP5
in \cite{Fergusson:2010gn}. From this constraint, we obtained constraints on 
model parameters. We emphasized that the amplitude of the trispectrum
for a particular configuration such as the regular tetrahedron limit
cannot be reliably used to characterise the amplitude of the
trispectrum and we need to use all configurations to define the amplitude
$g_{\rm NL} ^{equil}$. 

Finally, it is known that that inflation models based on
Galileon and its generalisations, which are shown to be
the most general single field inflation model
with second-order field equations
\cite{Kobayashi:2011nu,Charmousis:2011bf}
give also the equilateral type bispectrum
\cite{Mizuno:2010ag,Burrage:2010cu,Creminelli:2010qf,Kobayashi:2011pc,RenauxPetel:2011dv,RenauxPetel:2011uk,Gao:2011qe,DeFelice:2011uc}
(see also for the discussion about the shape dependence
of the bispectrum 
in this type of inflation model \cite{RenauxPetel:2011sb}). 
It would be interesting to study the amplitude of the trispectrum $g_{\rm NL} ^{equil}$
in these models. 

\begin{acknowledgments}
We would like to thank Rob Crittenden, Dominic Galliano
and Donough Regan for useful discussions. 
We also wish to thank Shinji Mukohyama and Takeshi Kobayashi for fruitful discussions.
We are grateful to Frederico Arroja and Maresuke Shiraishi for pointing typos
in numerical factors in Tables.
K.I. acknowledges supports by taiwan national science council under the project 
``detection of ultra-high energy cosmic neutrinos at south pole" 
and Japan-Russia Research Cooperative Program.
Part of this work was done during K.I. was supported by Grant-in-Aid for 
Scientific Research (A) No. 21244033. 
S.M. acknowledges support from the Labex P2IO of Orsay
and grateful to the ICG, Portsmouth
for their hospitality when this work was almost done.
K.K. is supported by the STFC (grant no. ST/H002774/1), a European Research Council Starting Grant and the Leverhulme trust.

\end{acknowledgments}

\appendix

\section{\label{appDBI}
Shape functions in general single field
k-inflation}

Here, based on our previous work \cite{Arroja:2009pd},
we summarise the shape functions for
the reduced trispectra in general single field
k-inflation models described by the following action:
\begin{equation}
S=\frac{1}{2}\int d^4x\sqrt{-g}\left[R+2P(X,\phi)
\right],\label{action}
\end{equation}
where $\phi$ is the inflaton field,
$R$  is the Ricci scalar and
$X\equiv-(1/2)g^{\mu\nu}\partial_\mu\phi\partial_\nu\phi$, where
$g_{\mu\nu}$ is the metric tensor.

In this class of models, the third and the fourth order
interaction Hamiltonian of the field perturbation
$\delta \phi$ in the flat gauge at leading order
in the slow-roll expansion are given by
\begin{eqnarray}
H_I^{(3)}(\eta)&=&\int
 d^3x\left[Aa\delta\phi'^3+Ba\delta\phi'
\left(\partial\delta\phi\right)^2\right]\,,
\label{k_inf_third_Hamiltonian}\\
H_I^{(4)}(\eta)&=&\int d^3x\left[\beta_1\delta\phi'^4+
\beta_2\delta\phi'^2\left(\partial\delta\phi\right)^2
+\beta_3\left(\partial\delta\phi\right)^4\right]\,,
\label{k_inf_fourth_Hamiltonian}
\end{eqnarray}
where prime denotes derivative with respect to conformal
time $\eta$ and coefficients $A$, $B$, $\beta_1$, $\beta_2$
and $\beta_3$ are given by
\begin{equation}
A=-\frac{\sqrt{2X}}{2}\left(P_{,XX}+\frac{2}{3} X P_{,XXX}\right), \quad B=\frac{\sqrt{2X}}{2}P_{,XX}.
\end{equation}
\begin{eqnarray}
\beta_1&=&P_{,XX}\left(1-\frac{9}{8}c_s^2\right)-2X
 P_{,XXX}\left(1-\frac{3}{4}c_s^2\right)+\frac{X^3
 c_s^2}{P_{,X}}P_{,XXX}^2-\frac{1}{6} X^2 P_{,4X},
\nonumber\\
\beta_2&=&-\frac{1}{2}P_{,XX}\left(1-\frac{3}{2}c_s^2\right)+\frac{1}{2}X
 c_s^2P_{,XXX},
\nonumber\\
\beta_3&=&-\frac{c_s^2}{8}P_{,XX}.
\end{eqnarray}

Here, $P_{,X}$ denotes the derivative of $P$ with respect to
$X$, $P_{,XX}$ denotes the second derivative of $P$
with respect to $X$, and so on. $c_s$ is the sound speed
of the perturbation of the scalar field $\phi$
which is defined as
\begin{eqnarray}
c_s^2 \equiv \frac{P_{,X}}{P_{,X} + 2 X P_{,XX}}.
\end{eqnarray}

The shape function
$S^{k} _{T}$ is composed of two parts
\begin{eqnarray}
S^{k} _{T} =
S^{k (cont)} _{T} +
S^{k (scalar)} _{T}\,,
\end{eqnarray}
where $S^{k (cont)} _{T}$
denotes the contribution from the contact interaction
and $S^{k (scalar)} _{T}$
denotes that from the scalar exchange interaction,
respectively.

The shape function for the reduced trispectrum arising from the contact interaction 
$S^{k (cont)} _\mathcal{T}$ depend on five parameters $k_1$, $k_2$, $k_3$,
$k_4$ and $k_{12}$. It is given by
\begin{eqnarray}
S^{k (cont)} _\mathcal{T} =
\left(-24 \beta_1 c_s ^3 S^{c_{1}} _\mathcal{T}
- \beta_2 c_s S^{c_{2}} _\mathcal{T}
- 2 \beta_3 c_s ^{-1} S^{c_{3}} _\mathcal{T} \right)
\frac{H^4}{4 X^2} N^8\,.
\label{shape_general_k_cont}
\end{eqnarray}
Here $S^{c_{1}} _\mathcal{T}$, $S^{c_{2}} _\mathcal{T}$
and $S^{c_{3}} _\mathcal{T}$ are the following
shape functions:
\begin{eqnarray}
S^{c_{1}} _\mathcal{T}
&=&
\frac{k_{12} \Pi_{i=1}^4k_i}{\left(\sum_{i=1}^4k_i\right)^5}
+ 3\;\; {\rm perms.}\,, \\
\label{shape_c1}
S^{c_{2}} _\mathcal{T}
 &=& \Biggl[\frac{k_{12}
 k_1^2k_2^2(\mathbf{k_3}\cdot\mathbf{k_4})}{\left(\sum_{i=1}^4k_i\right)^3\Pi_{i=1}^4k_i}\left(1+3\frac{(k_3+k_4)}{\sum_{i=1}^4k_i}+12\frac{k_3k_4}{\left(\sum_{i=1}^4k_i\right)^2}\right)
\nonumber\\
&&+\frac{k_{12} k_3^2k_4^2 (\mathbf{k_1}\cdot\mathbf{k_2})}
{\left(\sum_{i=1}^4k_i\right)^3\Pi_{i=1}^4k_i}\left(1+
3\frac{(k_1+k_2)}{\sum_{i=1}^4k_i}+
12\frac{k_1k_2}{\left(\sum_{i=1}^4k_i\right)^2}\right)
\Biggr]+3\;\; {\rm perms.}\,,
\label{shape_c2}\\
S^{c_{3}} _\mathcal{T}
&=& \frac{k_{12} (\mathbf{k_1}\cdot\mathbf{k_2})
(\mathbf{k_3}\cdot\mathbf{k_4})}
{\sum_{i=1}^4k_i\,\Pi_{i=1}^4k_i}
\left(1+\frac{\sum_{i<j}k_ik_j}{\left(\sum_{i=1}^4k_i\right)^2}+3\frac{\Pi_{i=1}^4k_i}{\left(\sum_{i=1}^4k_i\right)^3}\sum_{i=1}^4\frac{1}{k_i}+12\frac{\Pi_{i=1}^4k_i}{\left(\sum_{i=1}^4k_i\right)^4}\right)+3\;\; {\rm perms.}\,,
\label{shape_c3}
\end{eqnarray}
where ``$3\;\; {\rm perms.}$" denotes the permutations
$(k_1 \leftrightarrow k_2)$,
$(k_3 \leftrightarrow k_4)$ and $(k_1 \leftrightarrow k_2,k_3 \leftrightarrow k_4)$.
In Eq.~(\ref{shape_general_k_cont}),
$H$ is the Hubble parameter at inflation era and
$N=H/\sqrt{2 P_{,X} c_s}$.

Similarly, $S^{k (scalar)} _\mathcal{T}$
is given by
\begin{eqnarray}
S^{k (scalar)} _\mathcal{T} =
\left(A^2 c_s^4  S^{s_{1}} _\mathcal{T}
+ AB c_s^2 S^{s_{3}} _\mathcal{T}
+B^2 S^{s_{2}} _\mathcal{T} \right)
\frac{c_s^2 H^2 N^{10}}{8X^2}\,.
\label{shape_general_k_scalar}
\end{eqnarray}
Here $S^{s_{1}} _\mathcal{T}$, $S^{s_{2}} _\mathcal{T}$
and $S^{s_{3}} _\mathcal{T}$ are the following
shape functions:
\begin{eqnarray}
S^{s_1} _\mathcal{T}
 &=& -9 k_{12}
(k_1 k_2 k_3 k_4)^{1/2}
\biggl[ \tilde{\mathcal{F}}_1
(k_1,k_2,-k_{12},k_3,k_4,k_{12}) -
\tilde{\mathcal{F}}_1(-k_1,-k_2,-k_{12},k_3,k_4,k_{12})
\nonumber\\&&
+ \tilde{\mathcal{F}}_1
(k_3,k_4,-k_{12},k_1,k_2,k_{12}) -
\tilde{\mathcal{F}}_1(-k_3,-k_4,-k_{12},k_3,k_4,k_{12})
\biggr]+3\;\; {\rm perms.}\,,
\label{shape_s1}
\end{eqnarray}
\begin{eqnarray}
S^{s_{2}} _\mathcal{T} &=&
S^{s_{2a}} _\mathcal{T} +
S^{s_{2b}} _\mathcal{T} +
S^{s_{2c}} _\mathcal{T} +
S^{s_{2d}} _\mathcal{T}\,,
\label{def_shape_s2}\\
S^{s_{2a}} _\mathcal{T}
&=& -k_{12} (k_1 k_2 k_3 k_4)^{1/2}
(\mathbf{k_1}\cdot\mathbf{k_2})
(\mathbf{k_3}\cdot\mathbf{k_4})
\biggl[ \tilde{\mathcal{F}}_2
(-k_{12},k_1,k_2,k_{12},k_3,k_4) -
\tilde{\mathcal{F}}_2(-k_{12},-k_1,-k_2,k_{12},k_3,k_4)
\nonumber\\
&&+
 \tilde{\mathcal{F}}_2
(-k_{12},k_3,k_4,k_{12},k_1,k_2) -
\tilde{\mathcal{F}}_2(-k_{12},-k_3,-k_4,k_{12},k_1,k_2)
\biggr]+3\;\; {\rm perms.}\,,
\label{shape_s2a}
\\
S^{s_{2b}} _\mathcal{T}
 &=& -2 k_{12} (k_1 k_2 k_3 k_4)^{1/2}
 (\mathbf{k_1}\cdot\mathbf{k_2})
(\mathbf{k_{12}}\cdot\mathbf{k_4})
\biggl[ \tilde{\mathcal{F}}_2
(-k_{12},k_1,k_2,k_3,k_4,k_{12}) -
\tilde{\mathcal{F}}_2(-k_{12},-k_1,-k_2,k_3,k_4,k_{12})
\nonumber\\
&&+\tilde{\mathcal{F}}_2
(k_3,k_4,-k_{12},k_{12},k_1,k_2) -
\tilde{\mathcal{F}}_2(-k_3,-k_4,-k_{12},k_{12},k_1,k_2)
\biggr]+3\;\; {\rm perms.}\,,
\label{shape_s2b}
\\
S^{s_{2c}} _\mathcal{T}
 &=& 2 k_{12} (k_1 k_2 k_3 k_4)^{1/2}
 (\mathbf{k_{12}}\cdot\mathbf{k_2})
(\mathbf{k_3}\cdot\mathbf{k_4})
\biggl[ \tilde{\mathcal{F}}_2
(k_1,k_2,-k_{12},k_{12},k_3,k_4) -
\tilde{\mathcal{F}}_2(-k_1,-k_2,-k_{12},k_{12},k_3,k_4)
\nonumber\\
&&+\tilde{\mathcal{F}}_2
(-k_{12},k_3,k_4,k_1,k_2,k_{12}) -
\tilde{\mathcal{F}}_2(-k_{12},-k_3,-k_4,k_1,k_2,k_{12})
\biggr]+3\;\; {\rm perms.}\,,
\label{shape_s2c} \\
S^{s_{2d}} _\mathcal{T}
 &=& 4 k_{12} (k_1 k_2 k_3 k_4)^{1/2}
(\mathbf{k_{12}}\cdot\mathbf{k_2})
(\mathbf{k_{12}}\cdot\mathbf{k_4})
\biggl[ \tilde{\mathcal{F}}_2
(k_1,k_2,-k_{12},k_3,k_4,k_{12}) -
\tilde{\mathcal{F}}_2(-k_1,-k_2,-k_{12},k_3,k_4,k_{12})
\nonumber\\
&&+
\tilde{\mathcal{F}}_2
(k_3,k_4,-k_{12},k_1,k_2,k_{12}) -
\tilde{\mathcal{F}}_2(-k_3,-k_4,-k_{12},k_1,k_2,k_{12})
\biggr]+3\;\; {\rm perms.}\,,
\label{shape_s2d}
\end{eqnarray}
\begin{eqnarray}
S^{s_{3}} _\mathcal{T} &=&
S^{s_{3a}} _\mathcal{T} +
S^{s_{3b}} _\mathcal{T} +
S^{s_{3c}} _\mathcal{T} +
S^{s_{3d}} _\mathcal{T}\,,
\label{def_shape_s3}\\
S^{s_{3a}} _\mathcal{T}
 &=& 3 k_{12} (k_1 k_2 k_3 k_4)^{1/2}
 (\mathbf{k_3}\cdot\mathbf{k_4})
\biggl[ \tilde{\mathcal{F}}_3
(k_1,k_2,-k_{12},k_{12},k_3,k_4) -
\tilde{\mathcal{F}}_3(-k_1,-k_2,-k_{12},k_{12},k_3,k_4)
\nonumber\\
&&+\tilde{\mathcal{F}}_4
(-k_{12},k_3,k_4,k_1,k_2,k_{12}) -
\tilde{\mathcal{F}}_4(-k_{12},-k_3,-k_4,k_1,k_2,k_{12})
\biggr]+3\;\; {\rm perms.}\,,
\label{shape_s3a} \\
S^{s_{3b}} _\mathcal{T}
 &=& 6 k_{12} (k_1 k_2 k_3 k_4)^{1/2}
 (\mathbf{k_{12}}\cdot\mathbf{k_4})
\biggl[ \tilde{\mathcal{F}}_3
(k_1,k_2,-k_{12},k_3,k_4,k_{12}) -
\tilde{\mathcal{F}}_3(-k_1,-k_2,-k_{12},k_3,k_4,k_{12})
\nonumber\\
&&+\tilde{\mathcal{F}}_4
(k_3,k_4,-k_{12},k_1,k_2,k_{12}) -
\tilde{\mathcal{F}}_4(-k_3,-k_4,-k_{12},k_1,k_2,k_{12})
\biggr]+3\;\; {\rm perms.}\,,
\label{shape_s3b}\\
S^{s_{3c}} _\mathcal{T}
 &=& 3 k_{12} (k_1 k_2 k_3 k_4)^{1/2}
 (\mathbf{k_1}\cdot\mathbf{k_2})
\biggl[ \tilde{\mathcal{F}}_4
(-k_{12},k_1,k_2,k_3,k_4,k_{12}) -
\tilde{\mathcal{F}}_4(-k_{12},-k_1,-k_2,k_3,k_4,k_{12})
\nonumber\\
&&+\tilde{\mathcal{F}}_3
(k_3,k_4,-k_{12},k_{12},k_1,k_2) -
\tilde{\mathcal{F}}_3(-k_3,-k_4,-k_{12},k_{12},k_1,k_2)
\biggr]+3\;\; {\rm perms.}\,,
\label{shape_s3c}\\
S^{s_{3d}} _\mathcal{T}
 &=& -6 k_{12} (k_1 k_2 k_3 k_4)^{1/2}
 (\mathbf{k_{12}}\cdot\mathbf{k_2})
\biggl[ \tilde{\mathcal{F}}_4
(k_1,k_2,-k_{12},k_3,k_4,k_{12}) -
\tilde{\mathcal{F}}_4(-k_1,-k_2,-k_{12},k_3,k_4,k_{12})
\nonumber\\
&&+\tilde{\mathcal{F}}_3
(k_3,k_4,-k_{12},k_1,k_2,k_{12}) -
\tilde{\mathcal{F}}_3(-k_3,-k_4,-k_{12},k_1,k_2,k_{12})
\biggr]+3\;\; {\rm perms.} \,,
\label{shape_s3d}
\end{eqnarray}
where again ``$3\;\; {\rm perms.}$" denotes the permutations
$(k_1 \leftrightarrow k_3)$,
$(k_3 \leftrightarrow k_4)$ and $(k_1 \leftrightarrow k_2,k_3 \leftrightarrow k_4)$.
Here we have defined four $\tilde{\mathcal{F}}_i$
 functions (with $i=1,\ldots,4$) as follows;
\begin{eqnarray}
\tilde{\mathcal{F}}_1(k_1,k_2,k_3,k_4,k_5,k_6)
&=&-4|k_1 k_2 k_3 k_4 k_5 k_6|^\frac{1}{2}
\frac{1}{\mathcal{A}^3\mathcal{C}^3}\left(1+3\frac{\mathcal{A}}{\mathcal{C}}+6\frac{\mathcal{A}^2}{\mathcal{C}^2}\right)\,,\\
\tilde{\mathcal{F}}_2(k_1,k_2,k_3,k_4,k_5,k_6)
&=&-\frac{|k_1k_4|^\frac{1}{2}}{|k_2k_3k_5k_6|^\frac{3}{2}}\frac{1}{\mathcal{A}\mathcal{C}}
\bigg[
      1+\frac{k_5+k_6}{\mathcal{A}}+2\frac{k_5k_6}{\mathcal{A}^2}
      \nonumber\\&&
      +\frac{1}{\mathcal{C}}  \left(k_2+k_3+k_5+k_6+\frac{1}{\mathcal{A}}\left(\left(k_2+k_3\right)\left(k_5+k_6\right)+2k_5k_6\right)+2\frac{k_5k_6\left(k_2+k_3\right)}{\mathcal{A}^2}\right)
      \nonumber\\&&
      +\frac{2}{\mathcal{C}^2}\bigl(k_5k_6+\left(k_2+k_3\right)\left(k_5+k_6\right)+k_2k_3+\frac{1}{\mathcal{A}}\left(k_2k_3\left(k_5+k_6\right)+2k_5k_6\left(k_2+k_3\right)\right)\nonumber\\&&+2\frac{k_2k_3k_5k_6}{\mathcal{A}^2}\bigr)
 +\frac{6}{\mathcal{C}^3}\left(k_2k_3\left(k_5+k_6\right)+k_5k_6\left(k_2+k_3\right)+2\frac{k_2k_3k_5k_6}{\mathcal{A}}\right)
\nonumber\\
&&+24\frac{k_2k_3k_5k_6}{\mathcal{C}^4}
\bigg]\,,\\
\tilde{\mathcal{F}}_3(k_1,k_2,k_3,k_4,k_5,k_6)
&=&2\frac{|k_1k_2k_3k_4|^\frac{1}{2}}{|k_5k_6|^\frac{3}{2}}\frac{1}{\mathcal{A}\mathcal{C}^3}
\left[
      1+\frac{k_5+k_6}{\mathcal{A}}+2\frac{k_5k_6}{\mathcal{A}^2}
      +\frac{3}{\mathcal{C}}\left(k_5+k_6+2\frac{k_5k_6}{\mathcal{A}}\right)
      +12\frac{k_5k_6}{\mathcal{C}^2}
\right]\,,\\
\tilde{\mathcal{F}}_4(k_1,k_2,k_3,k_4,k_5,k_6)
&=&2\frac{|k_1k_4k_5k_6|^\frac{1}{2}}{|k_2k_3|^\frac{3}{2}}\frac{1}{\mathcal{A}^3\mathcal{C}}
\bigg[1+\frac{\mathcal{A}}{\mathcal{C}}+\frac{\mathcal{A}^2}{\mathcal{C}^2}+\frac{k_2+k_3}{\mathcal{C}}+2\frac{\mathcal{A}\left(k_2+k_3\right)+k_2k_3}{\mathcal{C}^2}
\nonumber\\&&
\qquad\qquad\qquad\qquad\qquad\quad
+3\frac{\mathcal{A}}{\mathcal{C}^3}\left(\mathcal{A}\left(k_2+k_3\right)+2k_2k_3\right)+12k_2k_3\frac{\mathcal{A}^2}{\mathcal{C}^4}\bigg]\,,
\end{eqnarray}
where $\mathcal{A}$ is defined by the sum of the last three arguments of
the $\tilde{\mathcal{F}}_i$ functions as  $\mathcal{A}=k_4+k_5+k_6$ and
$\mathcal{C}$ is defined by the sum of all the arguments as
$\mathcal{C}=k_1+k_2+k_3+k_4+k_5+k_6$.


To simplify the calculation, it is useful to notice that the following properties
\begin{eqnarray}
F(S^{c_1} _{\mathcal{T}},
\; S_{\mathcal{T}}) &=& \sum_i a_i
F(S^{c_1} _{\mathcal{T}},\;
S^{i} _{\mathcal{T}})\,,\\
F(S_{\mathcal{T}},
\; S_{\mathcal{T}})
&=& \sum_{i,j} a_i a_j
F(S^{i} _{\mathcal{T}},\;
S^{j} _{\mathcal{T}})\,,
\label{selfcorr_decomp}
\end{eqnarray}
hold for the shape function given by
\begin{eqnarray}
S_{\mathcal{T}}  = \sum_i a_i
S^{i} _{\mathcal{T}}\,,
\label{corr_decomp}
\end{eqnarray}
where $i=c_1, c_2, c_3, s_1, s_2, s_3$ and $a_i$'s
are corresponding coefficients.

Especially, in the case of single field DBI inflation,
the coefficients in the Hamiltonians
(\ref{k_inf_third_Hamiltonian}) and
(\ref{k_inf_fourth_Hamiltonian}) are given by
\begin{eqnarray}
&&A=-\frac{1}{2 \dot{\phi} c_s^5}\,,\;\;\;\;
B=\frac{1}{2 \dot{\phi} c_s^3}\,,\nonumber\\
&&\beta_1 = \frac{1}{2 c_s^7 \dot{\phi}^2}\,,\;\;\;\;
\beta_2=\frac{1}{4 c_s^3 \dot{\phi}^2}\,,\;\;\;\;
\beta_3 = -\frac{1}{8 c_s \dot{\phi}^2}\,,
\end{eqnarray}
and then the shape function based on the reduced trispectrum becomes
\begin{eqnarray}
S_\mathcal{T}^{DBI(\sigma)}=\frac{H^{12}}{{\dot \phi}^6 c_s^4}\left[
-3S_\mathcal{T}^{c1}+\frac{1}{64}S_\mathcal{T}^{s1}+\frac{1}{64}S_\mathcal{T}^{s2}-\frac{1}{64}S_\mathcal{T}^{s3}\right]\,.
\label{singleshapefunction}
\end{eqnarray}

In multi-field DBI inflation model \cite{Mizuno:2009mv},
in addition to the shape functions
$S^{c_{1}} _\mathcal{T}$, $S^{c_{2}} _\mathcal{T}$,
$S^{c_{3}} _\mathcal{T}$, $S^{s_{1}} _\mathcal{T}$
$S^{s_{2}} _\mathcal{T}$, $S^{s_{3}} _\mathcal{T}$,
we find it convenient to define
the following shape functions
$S^{\tilde{s}_{2}} _\mathcal{T}$ and
$S^{\tilde{s}_{3}} _\mathcal{T}$ given by
\begin{eqnarray}
S^{\tilde{s}_{2}} _\mathcal{T} &=&
S^{s_{2a}} _\mathcal{T} -
S^{s_{2b}} _\mathcal{T} -
S^{s_{2c}} _\mathcal{T} +
S^{s_{2d}} _\mathcal{T}\,,
\label{def_shape_s2_tilde}\\
S^{\tilde{s}_{3}} _\mathcal{T} &=&
S^{s_{3a}} _\mathcal{T} -
S^{s_{3b}} _\mathcal{T} +
S^{s_{3c}} _\mathcal{T} -
S^{s_{3d}} _\mathcal{T}\,.
\label{def_shape_s3_tilde}
\end{eqnarray}
With the above functions, the shape function  based on the reduced trispectrum
of the multi-field DBI inflation model can be expressed as
\begin{eqnarray}
S_\mathcal{T}^{DBI(s)}=\frac{H^{12}}{{\dot \phi}^6 c_s^4} 
T_{\cal RS} \left[
-\frac{1}{8}S_\mathcal{T}^{c2}+\frac{1}{576}S_\mathcal{T}^{s1}
+\frac{1}{64}S_\mathcal{T}^{\tilde s_2}
+\frac{1}{192}S_\mathcal{T}^{\tilde s_3} \right],
\label{multishapefunction}
\end{eqnarray}
where $T_{\cal RS}$ is the transfer coefficient
that relate the amplitude of original entropy perturbations
to the final curvature perturbation 
\footnote{Strictly speaking, there is another contribution to
the trispectrum in multi-field DBI inflation as pointed in
Ref.~\cite{RenauxPetel:2009sj}. 
Since the degree of the component studied in 
\cite{RenauxPetel:2009sj}
depends on background dynamics strongly, we concentrate
on the contribution coming from the intrinsically quantum
four-point function, for simplicity.}.

\section{Shape function in ghost inflation}
\label{appg}

In this appendix, based on our paper~\cite{Izumi:2010wm},
we review the trispectrum in ghost inflation and summarise its shape functions.
Ghost inflation is an inflation model where the inflation is driven by a
scalar field $\phi$ in the ghost condensation model.
The ghost condensation is the simplest Higgs phase for gravity in infrared
and in this model the four dimensional diffeomorphism is spontaneously broken by the timelike vacuum expectation value of the derivative of the scalar field $\phi$.
Thus, the action for the perturbative field $\pi$ of $\phi$ does not invariant under
four dimensional diffeomorphism and relevant terms  up to fourth order are
written generally as
\begin{eqnarray}
S=\int dt dx^3 a^3 \left[ \frac{1}{2} (\partial_t \pi)^2
- \frac{\alpha}{2 M^2}
\left(\frac{\vec{\nabla}^2}{a^2}\pi\right)^2
-  \frac{\beta}{2M^2} \partial_t \pi \frac{(\vec{\nabla} \pi)^2}{a^2}
-\frac{\gamma}{8M^4}
\frac{(\vec{\nabla} \pi)^4}{a^4}\right],
\label{eq:piaction}
\end{eqnarray}
where $\alpha$, $\beta$ and $\gamma$ are dimensionless constants of order unity and
$M$ is a constant with a dimension of mass.
Then, interaction Hamiltonian is written as
\begin{eqnarray}
H_I=\int dt d^3x  a^3\left[ \frac{\beta}{2M^2}\partial_t \pi
\frac{\left({\Vec \nabla}\pi\right)^2 }{a^2}+\frac{\tilde\gamma}{8M^4}
\frac{\left({\Vec \nabla}\pi\right)^4 }{a^4} \right],
\label{Hint_ghost}
\end{eqnarray}
where $\tilde \gamma =\gamma +2\beta^2$.

Trispectrum from the tree level contribution
can be decomposed into two parts.
One is obtained by the fourth order interaction Hamiltonian
and proportional to $\tilde \gamma$ which is
called the contact interaction contribution.
The other is obtained by the product of the third order interaction
Hamiltonian and independent of $\tilde \gamma$ which is
called the scalar exchange contribution.

Here, for simplicity we work only on the contact interaction contribution
of the trispectrum as only this contribution has new information
related with the four-point vertex. From Eq.~(\ref{Hint_ghost}),
if the condition $\gamma \gg \beta^2$  is satisfied,
this treatment can be justified.

According to \cite{Izumi:2010wm},
the shape function of the contact interaction contribution from the ghost inflation
based on the reduced trispectrum can be written as
\begin{eqnarray}
S_\mathcal{T}^{ghost}&\eq&
\frac{\tilde \gamma }{2^3 \alpha^{3/2}}\left(\frac{H}{M}\right)^9
 \left( \frac{\pi}{\Gamma(1/4)}\right)^4
\left( k_1 k_2 k_3 k_4 \right)^{1/2}k_{12}
({\bf k_1} \cdot {\bf k_2})({\bf k_3} \cdot {\bf k_4})
\nonumber\\
&&\qquad\qquad
\Re\biggl\{ i
\int^0_{-\infty}d\eta'
\left((-\eta')^{3/2}H_{3/4}^{(1)}(q_1 {\eta'}^2) \right)
\left((-\eta')^{3/2}H_{3/4}^{(1)}(q_2 {\eta'}^2) \right)
\nonumber\\
&&\qquad\qquad\qquad\qquad\qquad\qquad\qquad\qquad\qquad
\left((-\eta')^{3/2}H_{3/4}^{(1)}(q_3 {\eta'}^2) \right)
\left((-\eta')^{3/2}H_{3/4}^{(1)}(q_4 {\eta'}^2) \right)
\biggr\} ,
\label{ghostshapefunction}
\end{eqnarray}
where $q_i=\sqrt{\alpha}H k_i^2/(2M)$.
Here, we choose the reduced trispectrum such that depends on the 5 parameters
$k_1$, $k_2$, $k_3$, $k_4$ and $k_{12}$
as is the case in DBI inflation (see Appendix~\ref{appDBI}).

\section{Shape function in Lifshitz scalar}
\label{apph}

In this appendix, based on our previous paper~\cite{Izumi:2010yn},
we derive the shape function for the trispectrum in Lifshitz scalar.
Generally, some shapes of bispectra and trispectra
from the Lifshitz scalar are generated
because the action is not severely constrained by symmetries.
As we discuss in our previous paper~\cite{Izumi:2010yn}, however, only the local type bispectrum stems from the term which does not have the shift symmetry.
In this paper, since we want to distinguish between the models where
the equilateral type bispectrum are generated,
we concentrate on the model with the shift symmetry.
Moreover, only in the case with the dynamical critical exponent $z=3$
we can have the scale invariant power spectrum. Thus we choose $z=3$.

The terms up to fourth order of the action of Lifshitz scalar with $z=3$ and with the
shift symmetry in the ultra violet is written  as
\begin{eqnarray}
&& S = \frac{1}{2} \int dt d^3x\,  a(t)^3\biggl[ (\partial_t \phi)^2 +
\frac{1}{M^4 a(t)^6} \phi \Delta ^3 \phi +
\frac{1}{M^5 a(t)^6} \left\{ b_1  (\Delta^2 \phi) (\partial_i \phi)^2
 + b_2     (\Delta \phi)^3\right\} \nonumber\\
&&\qquad\qquad\qquad
 + \frac{1}{M^6 a(t)^6}
 \left\{ t_1\left( \Delta\phi \right)^2
\left(\partial_i \phi\right)^2+
t_2 \left( \partial_i\partial_j\phi \right)^2\left(\partial_k \phi\right)^2
+ t_3 (\partial_i \partial_j \partial_k \phi)
  (\partial_i \phi) (\partial_j \phi) (\partial_k \phi)\right\} \biggr],
\label{horavaaction}
\end{eqnarray}
where $M$, $b_1$, $b_2$, $t_1$, $t_2$ and $t_3$ are constants.
Then, the interaction Hamiltonian conforms with the non-linear terms of
the action given by Eq.~(\ref{horavaaction}).

We can decompose the contributions of the trispectrum into six parts which are
proportional to $b_1^2$, $b_1 b_2$, $b_2^2$, $t_1$, $t_2$ and $t_3$.
The scalar exchange contributions are the first three and the others are the contact
interaction contributions.
According to our previous paper~\cite{Izumi:2010yn},
the shape functions of these contributions can be written as
\begin{eqnarray}
&&S_\mathcal{T}^{h(se,ij)}=  \frac{ M^4}{2^3 \mu^4}\frac{
  (k_1^3 + k_2^3 + k_3^3 + k_4^3 + k_{12}^3)
  f_i(\boldsymbol{k_1}, \boldsymbol{k_2}) f_j(\boldsymbol{k_3},
 \boldsymbol{k_4})}{
 k_1 k_2 k_3 k_4 k_{12}^2 (k_1^3 + k_2^3+ k_{12}^3) (k_3^3 + k_4^3 +
 k_{12}^3) (k_1^3 + k_2^3 + k_3^3 + k_4^3) } ,
 \label{horavaseshapefunction}\\
&&S_\mathcal{T}^{h(ci,i)}= - \frac{ M^4}{2^3 \mu^4} \frac{ k_{12}
r_i(\boldsymbol{k_1}, \boldsymbol{k_2}, \boldsymbol{k_3},
 \boldsymbol{k_4})  }{k_1 k_2 k_3 k_4  (k_1^3 + k_2^3 +  k_3^3 +
 k_4^3)},
\label{Sci}\\
&&f_1(\boldsymbol{k_i}, \boldsymbol{k_j})
\equiv b_1 (k_i^6 + k_j^6 + k_{ij}^6 - k_i^2 k_j^4
  - k_i^4 k_j^2 - k_i^2 k_{ij}^4 - k_i^4 k_{ij}^2 - k_j^2 k_{ij}^4 -
  k_j^4 k_{ij}^2), \\
&&f_2(\boldsymbol{k_i}, \boldsymbol{k_j})
\equiv  6 b_2 k_i^2 k_j^2 k_{ij}^2,\\
&&r_1(\boldsymbol{k_i}, \boldsymbol{k_j}, \boldsymbol{k_k},\boldsymbol{k_l})=
4t_1 \left( k_i^2 k_j^2 (\boldsymbol{k_k}\cdot \boldsymbol{k_l})
	     + k_k^2 k_l^2(\boldsymbol{k_i}\cdot \boldsymbol{k_j}) \right), \\
&&r_2(\boldsymbol{k_i}, \boldsymbol{k_j}, \boldsymbol{k_k}, \boldsymbol{k_l})=
 4 t_2  \left(
  (\boldsymbol{k_i} \cdot \boldsymbol{k_j})^2(\boldsymbol{k_k} \cdot \boldsymbol{k_l})
 + (\boldsymbol{k_k} \cdot \boldsymbol{k_l})^2(\boldsymbol{k_i} \cdot \boldsymbol{k_j})
 \right) ,\\
&&r_3(\boldsymbol{k_i}, \boldsymbol{k_j}, \boldsymbol{k_k}, \boldsymbol{k_l})=
 2 t_3 \, \Bigl(
 (\boldsymbol{k_i} \cdot \boldsymbol{k_j})(\boldsymbol{k_i} \cdot
 \boldsymbol{k_k})(\boldsymbol{k_i} \cdot \boldsymbol{k_l})
 +  (\boldsymbol{k_j} \cdot \boldsymbol{k_i})(\boldsymbol{k_j} \cdot
 \boldsymbol{k_k})(\boldsymbol{k_j} \cdot \boldsymbol{k_l}) \nonumber\\
&& \qquad \qquad \qquad\qquad\qquad \qquad
 +  (\boldsymbol{k_k} \cdot \boldsymbol{k_i})(\boldsymbol{k_k} \cdot
 \boldsymbol{k_j})(\boldsymbol{k_k} \cdot \boldsymbol{k_l})
 +  (\boldsymbol{k_l} \cdot \boldsymbol{k_i})(\boldsymbol{k_l} \cdot
 \boldsymbol{k_j})(\boldsymbol{k_l} \cdot \boldsymbol{k_k})\Bigr),
\end{eqnarray}
where $S_\mathcal{T}^{h(se,ij)}$ and $S_\mathcal{T}^{h(ci,i)}$ are shape function propotional to
$b_i b_j$ and $t_i$, respectively.
Here, except for $S_\mathcal{T}^{h(ci,3)}$, we can choose the reduced trispectrum
such that it depends only on the 5 parameters $k_1$, $k_2$, $k_3$, $k_4$ and $k_{12}$
as is the case in DBI inflation
(see Appendix~\ref{appDBI}).
Only $S_\mathcal{T}^{h(ci,3)}$ cannot be reduced in such a manner because of
the presence of the product of $(\boldsymbol{k_1} \cdot \boldsymbol{k_2})$ and
$(\boldsymbol{k_1} \cdot \boldsymbol{k_3})$, etc.
Therefore, in the case of $S_\mathcal{T}^{h(ci,3)}$, we define the redused trispectrum
so that it is symmetric under the transpose among $\boldsymbol{k_1}$, $\boldsymbol{k_2}$,
$\boldsymbol{k_3}$ and $\boldsymbol{k_4}$.


\end{document}